\documentclass[twocolumn,nofootinbib,amsmath,amssymb,a4paper]{revtex4}

\usepackage{graphicx}
\usepackage{dcolumn}
\usepackage{bm}

\begin{document}

\begin{flushright}
\vspace*{-0.9truecm}
CERN-PH-TH/2007-017
\end{flushright}

\vspace*{-1.7truecm}

\title{\boldmath The $B\to\pi K$ Puzzle: A Status 
Report \unboldmath}

\author{Robert Fleischer}
 \email{Robert.Fleischer@cern.ch}
\affiliation{%
Theory Division, Physics Department, CERN, CH-1211 Geneva 23,
Switzerland
}%

\begin{abstract}
We discuss the theoretical interpretation of the $B \to\pi K$ system in 
the light of new data. Using the branching 
ratio and direct CP asymmetry of $B^0_d\to\pi^-K^+$, the picture of the direct 
CP violation in $B_d^0\to\pi^+\pi^-$ could be clarified: we predict 
${\cal A}_{\rm CP}^{\rm dir}(B_d\to\pi^+\pi^-)=-0.24\pm0.04$, which favours the
BaBar measurement, and extract $\gamma=\left(70.0^{+3.8}_{-4.3}\right)^\circ$, 
in agreement with the Standard-Model fits of the unitarity triangle. 
All $B\to\pi K$ modes with colour-suppressed electroweak penguin contributions
are found in excellent agreement with the Standard Model. The data for the
ratios $R_{\rm c,n}$ of the charged and neutral $B\to\pi K$ branching ratios,
which are sizeably affected by electroweak penguin contributions, have moved
quite a bit towards the Standard-Model predictions, which are almost unchanged,
thereby reducing the ``$B\to\pi K$ puzzle". On the other hand, the mixing-induced
CP violation of $B^0_d\to\pi^0K_{\rm S}$ still looks puzzling and could be 
accommodated through a modified electroweak penguin sector with a large
CP-violating new-physics phase, while the observed non-vanishing difference between 
the direct CP asymmetries of $B^\pm\to\pi^0K^\pm$ and $B_d\to\pi^\mp K^\pm$ 
seems to be caused by hadronic and not by new physics.
\end{abstract}

\maketitle

\section{Introduction}\label{sec:intro}
For more than a decade, the system of the $B\to\pi K$ decays is an outstanding 
topic in heavy-flavour physics (for a review, see \cite{RF-rev}). 
Thanks to the $B$ factories, we could obtain valuable insights into
these decays, raising the possibility of having a modified electroweak (EW) penguin
sector through the impact of new physics (NP). The following discussion follows
closely the strategy developed in \cite{BFRS}, and explores the 
picture after the experimental updates that were reported in the summer of 2006
\cite{BFRS-07}. The corresponding working assumptions for the treatment of
the hadronic sector can be summarised as follows: 
\begin{itemize}
\item[i)] {$SU(3)$ flavour symmetry:} however, $SU(3)$-breaking 
corrections are included through ratios of decay constants and form factors whenever they arise, and the sensitivity of the numerical results on non-factorizable 
$SU(3)$-breaking effects is explored.
\item[ii)] {Neglect of the penguin annihilation and exchange topologies:} these 
contributions can be probed and controlled through the $B_d\to K^+K^-$, 
$B_s\to\pi^+\pi^-$ system, which can be fully exploited at LHCb.
\end{itemize}
All consistency checks which can be performed with the current data support these
working assumptions and do not indicate any anomalous behaviour. Concerning 
the treatment of NP, we assume -- although we are basically performing 
a Standard-Model (SM) analysis -- that it manifests itself only in the electroweak (EW) 
penguin sector. Such a kind of physics beyond the SM can be accommodated, 
e.g., in SUSY, and models with extra $Z'$ bosons and extra dimension
scenarios. The topic of having NP in the EW penguin sector of $B\to\pi K$ decays 
has received a lot of attention in the literature (see, e.g., \cite{BpiK-papers}).

In the following discussion \cite{BFRS,BFRS-07}, we use the notation
\begin{eqnarray}
\lefteqn{\frac{\Gamma(B^0_d(t)\to f)-
\Gamma(\bar B^0_d(t)\to \bar f)}{\Gamma(B^0_d(t)\to f)+
\Gamma(\bar B^0_d(t)\to \bar f)}}\nonumber\\
&&={\cal A}_{\rm CP}^{\rm dir}\,\cos(\Delta M_d t)+
{\cal A}_{\rm CP}^{\rm mix}\,\sin(\Delta M_d t),\label{ACP-timedep}
\end{eqnarray}
where ${\cal A}_{\rm CP}^{\rm dir}$ and ${\cal A}_{\rm CP}^{\rm mix}$ denote the
``direct" and ``mixing-induced" CP-violating observables, respectively \cite{RF-rev};
a sign convention similar to that of (\ref{ACP-timedep}) will also be used for 
self-tagging $B_d$ and charged $B$ decays.

\boldmath
\section{The Starting Point: $B\to\pi\pi$}\label{sec:Bpipi}
\unboldmath
We have seen interesting progress in the exploration of CP violation in
$B^0_d\to\pi^+\pi^-$. In the SM, the decay amplitude of this decay can be
written as follows \cite{RF-Bpipi}:
\begin{equation}
A(B^0_d\to\pi^+\pi^-)=-|\tilde T| e^{i\delta_{\tilde T}}
\left[e^{i\gamma}-de^{i\theta}\right],
\end{equation}
where the $\tilde T$ amplitude is governed by the colour-allowed tree topologies, 
and the CP-conserving hadronic parameter $de^{i\theta}$ describes, sloppily 
speaking, the ratio of penguin to tree contributions. There is now -- for the 
first time -- a nice agreement between the BaBar and Belle measurements of the 
mixing-induced CP asymmetry:
\begin{equation}
{\cal A}_{\rm CP}^{\rm mix}(B_d\to\pi^+\pi^-)=
\left\{\begin{array}{ll}
0.53\pm0.14\pm0.02 & \mbox{(BaBar)}\\
0.61\pm0.10\pm0.04 & \mbox{(Belle),}
\end{array}\right.
\end{equation}
which yields the average of ${\cal A}_{\rm CP}^{\rm mix}(B_d\to\pi^+\pi^-)=
0.59\pm0.09$ \cite{HFAG}.
On the other hand, the picture of direct CP violation is still {\it not} settled:
\begin{equation}\label{ACP-dir-pipi-ex}
{\cal A}_{\rm CP}^{\rm dir}(B_d\to\pi^+\pi^-)=\left\{
\begin{array}{cc}
-0.16\pm0.11\pm0.03 & \mbox{(BaBar)}\\
-0.55\pm0.08\pm0.05 & \mbox{(Belle).}
\end{array}\right.
\end{equation}

This unsatisfactory situation can be resolved with the help of the $B^0_d\to\pi^-K^+$
mode, which is governed by QCD penguin contributions (this feature holds for all 
$B\to\pi K$ decays). Direct CP violation in this channel is now experimentally well 
established, with a nice agreement between the BaBar, Belle and CDF results, 
yielding an average of ${\cal A}_{\rm CP}^{\rm dir}(B_d\to\pi^\mp K^\pm)=
0.095\pm0.013$ \cite{punzi}. In the SM, the $B^0_d\to\pi^-K^+$ decay amplitude 
can be written as follows:
\begin{equation}
A(B^0_d\to \pi^- K^+)=P'\left[1-re^{i\delta}e^{i\gamma}\right].
\end{equation}
Using the $SU(3)$ flavour symmetry and the dynamical assumptions specified in 
Section~\ref{sec:intro}, we obtain 
\begin{equation}
re^{i\delta}=\frac{\epsilon}{d}e^{i(\pi-\theta)}
\end{equation}
with $\epsilon\equiv\lambda^2/(1-\lambda^2)=0.05$, implying the relation \cite{RF-Bpipi}:
\begin{eqnarray}
\lefteqn{H_{\rm BR}\equiv
\frac{1}{\epsilon}\left(\frac{f_K}{f_\pi}\right)^2\left[\frac{\mbox{BR}
(B_d\to\pi^+\pi^-)}{\mbox{BR}(B_d\to\pi^\mp K^\pm)}\right]}\nonumber\\
&&=-\frac{1}{\epsilon}\left[\frac{{\cal A}_{\rm CP}^{\rm dir}(B_d\to\pi^\mp 
K^\pm)}{{\cal A}_{\rm CP}^{\rm dir}(B_d\to\pi^+\pi^-)}\right].
\end{eqnarray}
Since the CP-averaged branching ratios and the direct CP asymmetry
${\cal A}_{\rm CP}^{\rm dir}(B_d\to\pi^\mp K^\pm)$ are well measured, we may use
this relation to {\it predict} the following value:
\begin{equation}
{\cal A}_{\rm CP}^{\rm dir}(B_d\to\pi^+\pi^-)=-0.24\pm0.04,
\end{equation}
which favours the BaBar result in (\ref{ACP-dir-pipi-ex}). Since we can express 
$H_{\rm BR}$, ${\cal A}_{\rm CP}^{\rm dir}(B_d\to\pi^\mp K^\pm)$ and 
${\cal A}_{\rm CP}^{\rm mix}(B_d\to\pi^+\pi^-)$ in terms of $\gamma$ and 
$d$, $\theta$, these parameters can be extracted from the data:
\begin{equation}\label{gam-d-theta}
\gamma=\left(70.0^{+3.8}_{-4.3}\right)^\circ,\quad
d=0.46\pm0.02,\quad \theta=(155\pm4)^\circ.
\end{equation}
The value of $\gamma$ is in agreement with the SM fits of the unitarity
triangle, and will be used for the remainder of this analysis. 

Applying the isospin symmetry, we may write
\begin{equation}
\begin{array}{rcl}
\sqrt{2}A(B^0_d\to\pi^0\pi^0)&=&|P|e^{i\delta_P}
\left[1+(x/d)e^{i\gamma}e^{i(\Delta-\theta)}\right]\\
\sqrt{2}A(B^+\to\pi^+\pi^0)&=&-|\tilde T|e^{i\delta_{\tilde T}}e^{i\gamma}
\left[1+xe^{i\Delta}\right],
\end{array}
\end{equation}
where the hadronic parameter $xe^{i\Delta}$ denotes the ratio of 
``colour-suppressed" to ``colour-allowed tree'' amplitudes. The experimental
values of the ratios of the CP-averaged $B\to\pi\pi$ branching ratios allow
an extraction of this quantity, with the following result:
\begin{equation}
x=0.92_{-0.09}^{+0.08},\quad \Delta=-(50_{-14}^{+11})^\circ.
\end{equation}
Complementing these numbers with those in (\ref{gam-d-theta}), the
following {\it predictions} can be made in the SM:
\begin{eqnarray}
{\cal A}_{\rm CP}^{\rm dir}(B_d\to\pi^0\pi^0)
&=&-(0.40^{+0.14}_{-0.21})\label{ACPdir00}\\
{\cal A}_{\rm CP}^{\rm mix}(B_d\to\pi^0\pi^0)&=&
-(0.71^{+0.16}_{-0.17}),
\end{eqnarray}
which offer the exciting perspective of observing {\it large} CP violation in the
$B^0_d\to\pi^0\pi^0$ channel. So far, only data for the direct CP asymmetry
are available from the BaBar and Belle collaborations, with the average of 
${\cal A}_{\rm CP}^{\rm dir}(B_d\to\pi^0\pi^0)=-(0.36^{+0.33}_{-0.31})$, which is --
note the signs -- in remarkable agreement with (\ref{ACPdir00}), giving us
further confidence in our analysis.

\boldmath
\section{The Main Target: $B\to\pi K$}\label{sec:BpiK}
\unboldmath
The $B\to\pi K$ decays are dominated by QCD penguin topologies, and can
be divided into two classes, depending on the impact of EW penguins:
\begin{itemize}
\item The EW penguins are colour-suppressed, leading to tiny contributions:
$B^0_d\to\pi^-K^+$, $B^+\to\pi^+K^0$.
\item The EW penguins are colour-allowed, leading to sizeable effects:
$B^0_d\to\pi^0K^0$, $B^+\to\pi^0K^+$.
\end{itemize}

\subsection{Observables with Tiny EW Penguin Effects}
Let us first have a closer look at the $B\to\pi K$ observables with a tiny
impact of the EW penguins. For the determination of $\gamma$ 
discussed above, we have already used the CP-averaged branching ratio 
and the direct CP asymmetry of $B^0_d\to\pi^- K^+$, yielding a value of $\gamma$
in excellent agreement with the SM fits of the unitarity triangle. 
Another decay with colour-suppressed EW penguins is at our disposal, with
the following amplitude:
\begin{equation}
A(B^+\to\pi^+K^0)=-P'\left[1+\rho_{\rm c}e^{i\theta_{\rm c}}
e^{i\gamma}\right],
\end{equation}
where the doubly Cabibbo-suppressed parameter $\rho_{\rm c}e^{i\theta_{\rm c}}$ is 
usually neglected, implying vanishing direct CP violation. This feature is nicely
supported by the experimental average 
${\cal A}_{\rm CP}^{\rm dir}(B^\pm\to\pi^\pm K)=-0.009\pm0.025$ \cite{HFAG}. 

Finally, the working assumptions specified in Section~\ref{sec:intro}
allow us to predict the following ratio:
\begin{eqnarray}
\lefteqn{R\equiv \left[\frac{\mbox{BR}(B_d\to\pi^\mp K^\pm)}{\mbox{BR}(B^\pm
\to\pi^\pm K)}\right]\frac{\tau_{B^+}}{\tau_{B^0_d}}}\nonumber\\
&&\stackrel{\rm SM}{=}0.942\pm0.012 
\stackrel{\rm exp}{=}0.93\pm0.05.
\end{eqnarray}
Consequently, we obtain an excellent agreement with the SM, and no anomalous 
value of $\rho_{\rm c}$ is indicated,\footnote{This picture of $\rho_{\rm c}$ follows 
also from $B^\pm\to K^\pm K$ decays \cite{FR}.} thereby ruling out toy models of 
final-state interaction effects that were discussed several years ago. 

The strategy developed in \cite{BFRS} allows also the prediction of the observables 
of the $B_s\to K^+K^-$ decay, where the impact of EW penguins is tiny 
(colour-suppressed) as well. In the SM, the corresponding CP asymmetries are 
predicted as follows:
\begin{eqnarray}
{\cal A}_{\rm CP}^{\rm dir}(B_s\to K^+K^-)&=&
0.093\pm0.015\\
{\cal A}_{\rm CP}^{\rm mix}(B_s\to K^+K^-)&=&-0.234_{-0.014}^{+0.017}.
\end{eqnarray}
In contrast to the CP asymmetries, an $SU(3)$-breaking form-factor
ratio enters the prediction of the CP-averaged branching ratio. Using
the result of a recent QCD sum-rule calculation \cite{Khod} yields
\begin{equation}\label{BsKK-BR}
\mbox{BR}(B_s\to K^+K^-)=\left\{
\begin{array}{ll}
(27.9_{-5.1}^{+7.1})\times 10^{-6} & \mbox{[$B\to\pi\pi$]}\\
(28.1_{-5.1}^{+7.0})\times 10^{-6} & \mbox{[$B\to\pi K$].}
\end{array}
\right.
\end{equation}
As indicated, there are two options for the prediction of this branching ratio, using 
either $B\to\pi\pi$ or $B\to\pi K$ data, which are in remarkable agreement with each
other. The $B_s\to K^+K^-$ channel has recently been observed at CDF,
with the following branching ratio \cite{punzi}:
\begin{equation}\label{BsKK-exp}
\mbox{BR}(B_s\to K^+K^-)=(24.4\pm1.4\pm4.6)\times10^{-6}.
\end{equation}
Within the uncertainties, (\ref{BsKK-BR}) is in nice agreement with (\ref{BsKK-exp}),
which is another support of the assumptions listed in Section~\ref{sec:intro}.
The $B_s\to K^+K^-$, $B_d\to\pi^+\pi^-$ system offers a powerful $U$-spin
strategy for the extraction of $\gamma$ at LHCb \cite{RF-Bpipi,nardulli}; the 
predictions and hadronic parameters given above are useful for further 
experimental studies to prepare the real data taking at the LHC. 

\subsection{Observables with Sizeable EW Penguin Effects}
The following ratios are key quantities for an analysis of the $B\to\pi K$ system:
\begin{eqnarray}
R_{\rm c}&\equiv&2\left[
\frac{\mbox{BR}(B^\pm\to\pi^0K^\pm)}{\mbox{BR}(B^\pm\to\pi^\pm K^0)}\right]
\stackrel{\rm exp}{=}1.11\pm0.07\\
R_{\rm n}&\equiv&
\frac{1}{2}\left[
\frac{\mbox{BR}(B_d\to\pi^\mp K^\pm)}{\mbox{BR}(B_d\to\pi^0K^0)}\right]
\stackrel{\rm exp}{=}0.99\pm0.07.
\end{eqnarray}
The EW penguins, which provide an interesting avenue for NP to 
manifest itself \cite{EWP-NP}, enter here in colour-allowed form through 
the modes involving neutral pions, and are theoretically described by two 
parameters: $q$, which measures the ``strength" of the EW penguin with 
respect to the tree contributions, and a CP-violating phase $\phi$. In the SM, 
the $SU(3)$ flavour symmetry allows a prediction of $q=0.60$ \cite{NR}, and $\phi$ 
{\it vanishes.}

\begin{figure}
\includegraphics[width=0.42\textwidth]{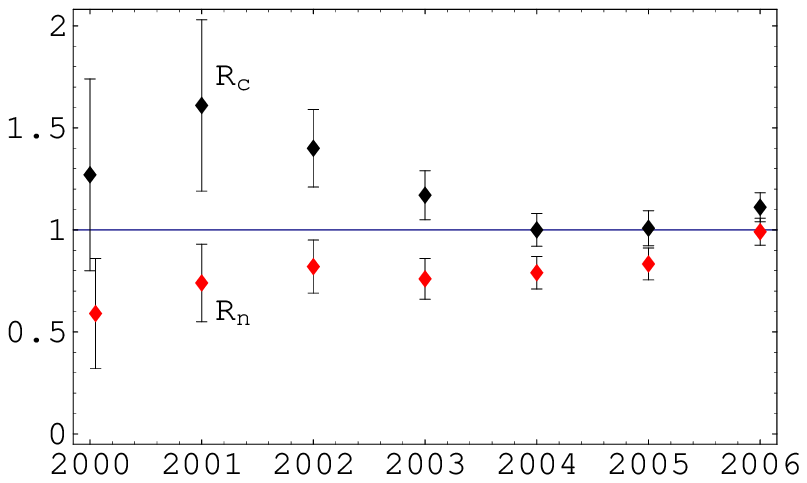}
\caption{\label{fig:timeline} The time evolution of the experimental values of 
$R_{\rm c,n}$.}
\end{figure}

\begin{figure}[t]
\includegraphics[width=0.42\textwidth]{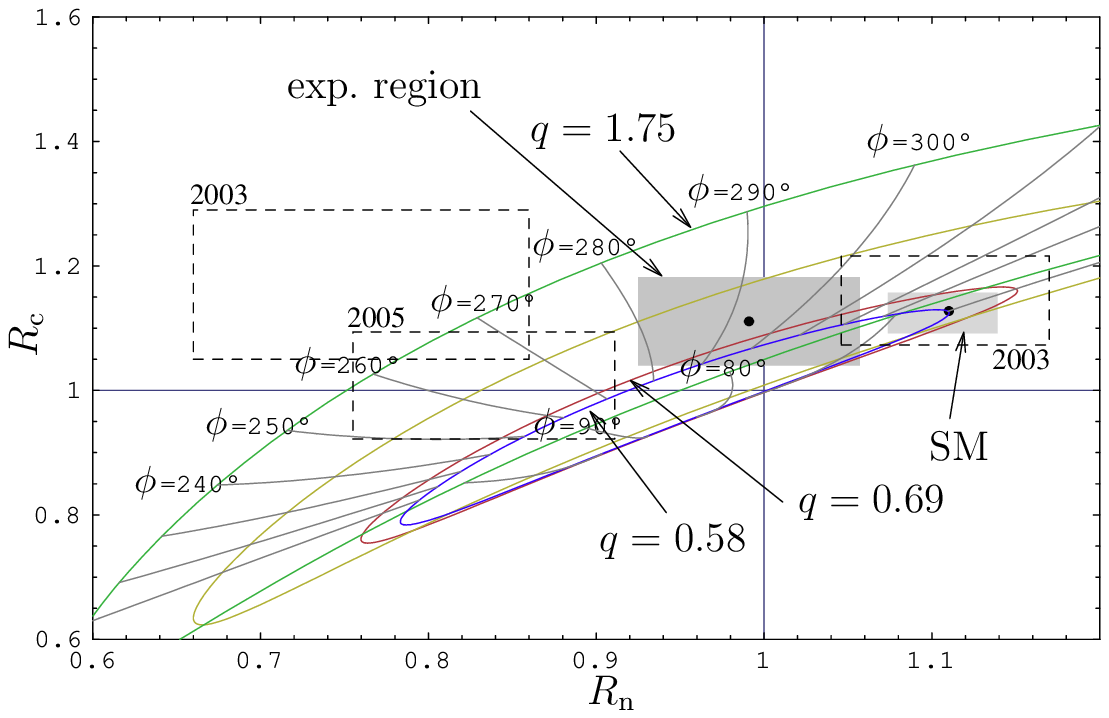}
\caption{\label{fig:RnRc} The situation in the $R_{\rm n}$--$R_{\rm c}$ plane.}
\end{figure}

If we look at Fig.~\ref{fig:timeline} showing the time evolution of the experimental 
values of $R_{\rm c}$ and $R_{\rm n}$, we observe that the central values have significantly moved up (partly due to radiative corrections affecting final states with 
charged particles \cite{BarIsi}), while the errors were only marginally reduced.
In Fig.~\ref{fig:RnRc}, we show the situation in the plane of the observables
$R_{\rm n}$ and $R_{\rm c}$: the contours correspond to different values of
$q$, and are parametrized through the phase $\phi$. We see that the SM
prediction (on the right-hand side) is very stable in time, having now significantly
reduced errors. On the other hand, the $B$-factory data have moved quite
a bit towards the SM. Converting the experimental values of
$R_{\rm n}$ and $R_{\rm c}$ into $q$ and $\phi$ yields
\begin{equation}
q = 0.65_{-0.35}^{+0.39},\quad
\phi = -(52^{+21}_{-50})^\circ. 
\end{equation}
A similar trend -- see, in particular, the time evolution of
$(\sin2\beta)_{\phi K_{\rm S}}$ -- is also present in the analysis of 
CP violation in $b\to s$ penguin-dominated decays \cite{HFAG}. 

\begin{figure}
\vspace*{0.5truecm}
\includegraphics[width=0.42\textwidth]{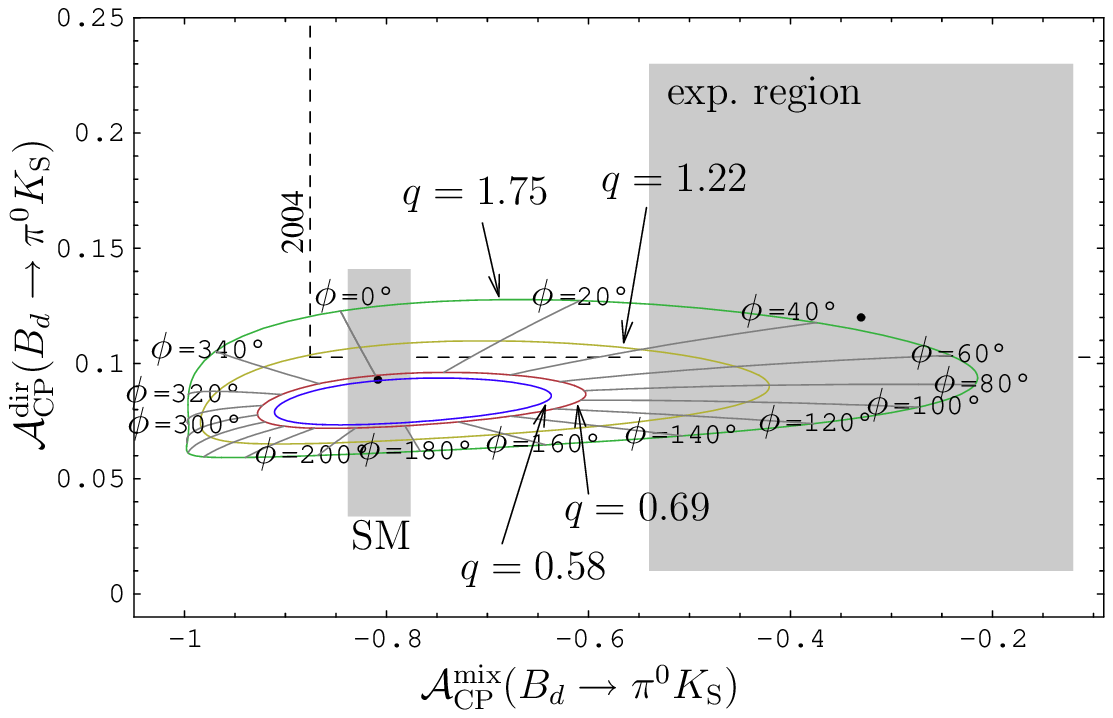}
\caption{\label{fig:ACP}The
${\cal A}_{\rm CP}^{\rm mix}(B_d\to\pi^0K_{\rm S})$--${\cal A}_{\rm CP}^{\rm 
dir}(B_d\to\pi^0K_{\rm S})$ plane.}
\end{figure}

Let us now have a closer look at the CP asymmetries of the 
$B^0_d\to\pi^0 K_{\rm S}$ and $B^\pm\to\pi^0K^\pm$ channels, which have
received a lot of attention and can also be analysed in the strategy of \cite{BFRS}.
As can be seen in Fig.~\ref{fig:ACP}, SM predictions for the CP-violating observables
of $B^0_d\to\pi^0K_{\rm S}$ are obtained that are much sharper than the current
$B$-factory data. In particular ${\cal A}_{\rm CP}^{\rm mix}(B_d\to\pi^0K_{\rm S})$
offers a very interesting quantitiy. We also see that the experimental central
values can be reached for large {\it positive} values of $\phi$. Concerning
direct CP violation in $B^\pm\to\pi^0K^\pm$, the following SM prediction
arises:
\begin{equation}
{\cal A}^{\rm dir}_{\rm CP}(B^\pm\to\pi^0K^\pm)=
-0.001^{+0.049}_{-0.041},
\end{equation}
which is in good agreement with the experimental result $-0.047\pm0.026$
within the errors. For the new input data, this feature turns interestingly out to 
be almost independent of NP. Consequently, the non-vanishing experimental
value of 
\begin{eqnarray}
\Delta A&\equiv& {\cal A}_{\rm CP}^{\rm dir}(B^\pm\to\pi^0K^\pm)-
{\cal A}_{\rm CP}^{\rm dir}(B_d\to\pi^\mp K^\pm)\nonumber\\
&\stackrel{\rm exp}{=}&-0.140\pm0.030,
\end{eqnarray}
which differs from zero at the $4.7\,\sigma$ level, is likely to be generated through hadronic effects, i.e.\ not through the presence of NP. 

Performing, finally, a fit to $R_{\rm n}$, $R_{\rm c}$ and the CP asymmetries
of $B^0_d\to\pi^0K_{\rm S}$ yields
\begin{equation}
q=1.7_{-1.3}^{+0.5},\quad \phi=+\left(73_{-18}^{+6}\right)^\circ.
\end{equation}
Interestingly, these parameters -- in particular the large {\it positive} phase -- 
would also allow us to accommodate the experimental values of
$(\sin2\beta)_{\phi K_{\rm S}}$ and the CP asymmetries of other 
$b\to s$ penguin modes with central values smaller than 
$(\sin2\beta)_{\psi K_{\rm S}}$. The large value of $q$ would be excluded 
by constraints from rare decays in simple scenarios where NP enters only 
through $Z$ penguins \cite{BFRS}, but could still be accommodated in other 
scenarios, e.g.\ in models with leptophobic $Z'$ bosons.

\section{Conclusions}\label{sec:concl}
The strategy developed in \cite{BFRS} continues to provide a powerful tool
for the theoretical interpretation of the $B\to\pi\pi,\pi K$ data \cite{BFRS-07}.
Thanks to the progress at the $B$ factories, now only data could be used
where the BaBar and Belle collaborations are in full agreement with each other. 
However, the corresponding SM predictions are very stable, with almost 
unchanged central values since the original analysis of 2003, and significantly 
reduced errors. 

Using the braching ratio and direct CP asymmetry of the 
$B^0_d\to\pi^-K^+$ channel, the picture of direct CP violation 
in $B^0_d\to\pi^+\pi^-$ could be clarified, with the prediction of
${\cal A}_{\rm CP}^{\rm dir}(B_d\to\pi^+\pi^-)=-0.24\pm0.04$, which
favours the BaBar result, and the extraction of 
$\gamma=\left(70.0^{+3.8}_{-4.3}\right)^\circ$, which is in agreement with 
the SM fits of the unitarity triangle. 

The current status of the $B\to\pi K$ system can be summarized 
as follows:
\begin{itemize}
\item All modes with colour-suppressed EW penguins are in excellent 
agreement with the SM.

\item The data for the $R_{\rm n,c}$ have moved quite a bit towards the SM predictions,
which are almost unchanged, thereby reducing the ``$B\to\pi K$ puzzle" for the
CP-averaged branching ratios. 

\vspace*{0.2truecm}

\item The non-zero experimental value of $\Delta A$ seems 
to be caused by hadronic and not by NP effects. 

\item On the other hand, the mixing-induced CP violation in $B^0_d\to\pi^0K_{\rm S}$ 
still looks puzzling, and can straightforwardly be accommodated through a modified 
EW penguin sector with a large, positive value of the CP-violating NP phase $\phi$.
\end{itemize}
Unfortunately, we still cannot draw definite conclusions about the presence of
NP in the $B\to\pi K$ system (and other $b\to s$ penguin decays, such as
$B^0_d\to\phi K_{\rm S}$). It will be interesting to keep track of the picture
of these decays once the data improve further.

\begin{acknowledgments}
I would like to thank Giovanni Punzi for valueable discussions about the
most recent CDF results.
\end{acknowledgments}

\end{document}